%% file: main.tex
\documentclass[runningheads]{llncs}
\input{preamble}

\newif\ifanonymous

\title{Agentic-IC3: Enabling Semantic Proof Search in IC3 Model Checking}
\titlerunning{Agentic-IC3: Enabling Semantic Proof Search in IC3 Model Checking}
\hypersetup{pdftitle={Agentic-IC3: Enabling Semantic Proof Search in IC3},bookmarksdepth=3}
\ifanonymous
  \author{Anonymous Author(s)}
  \authorrunning{Anonymous Author(s)}
  \institute{}
  \hypersetup{pdfauthor={}}
\else
  \author{Yu-Wei Fan \and SooHyuk Cho \and Aarti Gupta \and Sharad Malik}
  \authorrunning{Y.-W. Fan et al.}
  \institute{Princeton University\\
    \email{\{yf9172,soohyuk.cho,sharad\}@princeton.edu}\\
    \email{aartig@cs.princeton.edu}}
  \hypersetup{pdfauthor={Yu-Wei Fan, SooHyuk Cho, Aarti Gupta, Sharad Malik}}
\fi

\begin{document}
\maketitle
\input{sections/abstract}
\input{sections/introduction}
\input{sections/background}
\input{sections/motivation}
\input{sections/framework}
\input{sections/evaluation}
\input{sections/discussion}
\input{sections/related-work}
\input{sections/conclusion}

\bibliographystyle{splncs04}
\bibliography{references}

\clearpage
\end{document}

%% file: preamble.tex
\usepackage[utf8]{inputenc} 
\usepackage[T1]{fontenc}    
\usepackage{lmodern}        
\usepackage{url}            
\usepackage{booktabs}       
\usepackage{amsfonts}       
\usepackage{nicefrac}       
\usepackage{microtype}      
\usepackage{xcolor}         
\usepackage{graphicx}
\usepackage{multirow}
\usepackage{listings}
\usepackage{float}

\usepackage{algorithm}
\usepackage[noend]{algpseudocode}
\usepackage[hidelinks]{hyperref}       
\usepackage{cleveref}
\AtBeginDocument{}
\crefname{algline}{line}{lines}
\Crefname{algline}{Line}{Lines}
\makeatletter
\providecommand{\theHALG@line}{}
\renewcommand{\theHALG@line}{\thealgorithm.\arabic{ALG@line}}
\newcommand{\alglabel}[1]{%
  \begingroup
  \edef\cref@currentlabel{[algline][\arabic{ALG@line}][]\arabic{ALG@line}}%
  \label{#1}%
  \endgroup}
\makeatother

\lstdefinestyle{rtl}{
  language=Verilog,
  basicstyle=\small\ttfamily,
  breaklines=true,
  keywordstyle=\color{blue!65!black}\bfseries,
  commentstyle=\color{green!40!black}\itshape,
  stringstyle=\color{orange!70!black},
  morekeywords={wire,reg,localparam,assume,assert,property,endproperty},
  columns=fullflexible,
  keepspaces=true,
  showstringspaces=false,
  frame=single,
  rulecolor=\color{black!25},
  aboveskip=0pt,
  belowskip=0pt
}

\newif\ifdraft 
\drafttrue

\newif\ifrevisiondiff
\revisiondifffalse

%% file: sections/abstract.tex
\begin{abstract}
IC3 is a state-of-the-art algorithm for hardware model checking that proves safety properties by incrementally constructing an inductive invariant consisting of a set of lemmas. 
Its effectiveness depends on generalization heuristics that identify useful lemmas and guide proof search. 
However, many leading IC3 hardware model checkers operate on lowered, bit-level representations, where high-level design relationships are difficult to exploit for generalization.
Those operating at a higher level remain limited in exploiting high-level design structure and semantics. 
We present Agentic-IC3, built on Pono's word-level model-checking infrastructure, which integrates a language-model agent into IC3 to guide semantic proof search using register-transfer-level (RTL) design information. 
The framework exposes an agent-oriented interface to a persistent IC3 backend, allowing the agent to interact with an explicit, evolving proof state throughout verification.
Across successive proof obligations, the agent relates intermediate proof states and solver feedback to the RTL and proposes high-level lemmas through both SAT and UNSAT generalization. 
Beyond generalization, the agent can introduce derived observation signals to express design relationships succinctly and obtain more informative feedback, and backtrack to revise proposals that lead to unproductive proof branches. 
The backend checks proposals before updating the proof state, preserving soundness and providing feedback for further reasoning. 
On a suite of 14 benchmarks spanning security information-flow verification and functional verification of communication protocols, processors, and functional units, 
Agentic-IC3 solves 10 cases within a one-hour timeout, including four unsolved by all three evaluated baselines: rIC3, Pono-IC3Bits, and $\mathcal{A}$-IC3.
\end{abstract}

%% file: sections/introduction.tex
\section{Introduction}
\label{sec:introduction}

Hardware model checking is a widely used approach to hardware formal verification.
Given a hardware design and a property, a model checker determines whether the property holds for all executions of the design.
IC3~\cite{Bradley2011}, also known as property-directed reachability (PDR)~\cite{PDR}, is a state-of-the-art algorithm for checking safety properties, which require that an undesirable condition never occurs.
It constructs an inductive invariant: a collection of facts, called \emph{lemmas}, that holds initially, is preserved by each transition, and implies the property.

IC3 combines proof-search guidance with solver-based checks.
Human-designed heuristics determine how proof obligations are generalized and which candidate lemmas the checker attempts to prove, while solvers check the validity of each proof step.
A central task is generalization: turning information about concrete states into lemmas that cover broader sets of states and help construct the proof~\cite{BetterGeneralization,IGoodLemmas,PredictingLemmas,BCCIC3}.
Useful generalizations depend on the design behavior and the property being checked.
A rule useful in one application may provide little guidance in another, making a general strategy difficult to develop.

The design of generalization heuristics also depends on how the model checker represents the hardware and its lemmas.
Bit-level IC3 operates on individual state bits, where word-level operations in the original design have been lowered into Boolean logic.
This restricted representation supports efficient generalization through operations such as literal removal, backed by optimized SAT solvers and compact data structures~\cite{PDR,IC3Variants}.
However, high-level relationships can be difficult to identify in the lowered design, and expressing them may require multiple bit-level lemmas.
Word-level IC3 retains operations over multi-bit values, allowing lemmas to express relationships such as equality or arithmetic constraints directly~\cite{IC3SA,WordLevelPDR,SyGuSAPDR}.
The richer representation offers more expressive lemmas, but searching the larger space requires more resources, and heuristics may miss useful candidates.

Developing effective generalization strategies therefore requires substantial expert effort.
Existing approaches encode useful proof patterns through heuristics, synthesis grammars, and abstraction rules~\cite{IC3Variants,IC3SA,SyGuSAPDR}.
Other work extends the proof vocabulary~\cite{PDRER} or exploits application-specific structure~\cite{SecIC3}.
Their effectiveness depends on how well the generalization rules and candidate representations capture relationships relevant to the target designs and properties, limiting how well the guidance transfers across applications.

ML-based approaches offer another way to obtain this guidance, drawing on data and verification feedback.
Within bit-level IC3, NeuroPDR and DeepIC3 use graph neural networks to guide inductive generalization~\cite{NeuroPDR,DeepIC3}, while $\mathcal{A}$-IC3 learns to select among generalization strategies during verification~\cite{AIC3}.
Recent work also uses language models to generate invariants from design information and solver feedback.
Large Lemma Miners and CIll propose helper invariants for hardware verification and validate them with formal tools~\cite{LargeLemmaMiners,CIll}, while IC3Syn uses language models for generalization in an IC3-based procedure for distributed protocols~\cite{IC3Syn}.
These efforts motivate a broader role for language-model reasoning within IC3: using evidence accumulated during proof search to propose lemmas and revise the decisions that determine subsequent proof obligations.

We present Agentic-IC3, a framework built on Pono's word-level model-checking infrastructure~\cite{Pono} that integrates a language-model agent into IC3 as a proof-search policy.
The interface deliberately keeps the agent's view of both the design and the proof state at the register-transfer level (RTL).
The agent examines the RTL, property, proof states, and solver feedback to identify signal relationships that explain an obligation's unreachability and express them as lemmas.

The backend maintains the ongoing proof state, including admitted lemmas and proof obligations.
Through an agent-oriented interface to a persistent backend session, the agent uses this evolving context to guide successive generalization steps.
The interaction also supports decisions beyond lemma generation.
The agent can introduce derived observation signals to make useful relationships easier to recognize and express, and backtrack to earlier proof obligations to revise proposals that lead to unproductive branches.
These actions let the agent shape both the representation and direction of the continuing proof search.
The backend validates proof updates with SAT/SMT solvers, preserving soundness and providing feedback for subsequent reasoning.

We evaluate Agentic-IC3 on 14 benchmarks with applications spanning security information-flow and functional verification of communication protocols, processors, and functional units.
With a one-hour timeout per benchmark, Agentic-IC3 solves 10 cases, including four unsolved by all three evaluated baselines: rIC3~\cite{rIC3}, Pono-IC3Bits~\cite{Pono}, and $\mathcal{A}$-IC3~\cite{AIC3}.
These results demonstrate the potential of combining model checking algorithms with agent-guided semantic reasoning on challenging hardware verification tasks.

%% file: sections/background.tex
\section{Background}
\label{sec:background}

We model a hardware design as a transition system with state variables $X$, an initial-state predicate $\mathrm{Init}(X)$, and a transition relation $T(X,X')$, where primed variables denote the next state.
A safety property $P(X)$ requires that every reachable state satisfy $P$; a state satisfying $\neg P$ is a bad state.
An inductive invariant $\mathrm{Inv}(X)$ proves the property when
\[
\mathrm{Init}\Rightarrow\mathrm{Inv},\qquad
\mathrm{Inv}\land T\Rightarrow\mathrm{Inv}',\qquad
\mathrm{Inv}\Rightarrow P.
\]

\begin{figure}[!tbp]
  \centering
  \includegraphics[width=.9\linewidth]{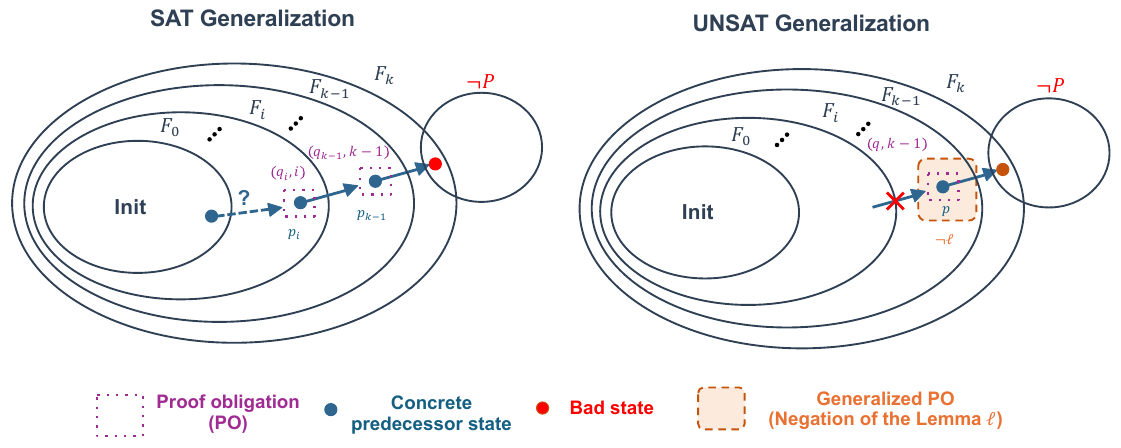}
  \caption{IC3 proof search and generalization.
  Left: backward predecessor search creates proof obligations, with SAT generalization expanding concrete predecessors into regions.
  The question mark denotes a pending predecessor query.
  Right: an UNSAT predecessor query enables a blocking lemma $\ell$ that excludes the larger region $\neg\ell$.
  }
  \label{fig:ic3-explanation}
\end{figure}

IC3 constructs $\mathrm{Inv}$ by maintaining a sequence of frames $F_0,\ldots,F_k$.
Each frame $F_i(X)$ is a formula over the state variables that overapproximates the states reachable within $i$ transitions, with $F_0=\mathrm{Init}$.
The nested ovals in Figure~\ref{fig:ic3-explanation} illustrate the inclusion between frames, which satisfy
\[
F_i\Rightarrow F_{i+1},\qquad
F_i\land T\Rightarrow F_{i+1}'.
\]
IC3 refines these overapproximations by adding lemmas to the frames.
A lemma in $F_i$ holds within the first $i$ transitions but may not hold for arbitrarily many transitions~\cite{Bradley2011}.

After checking that the property holds initially, IC3 searches for a bad state satisfying $F_k\land\neg P$.
Blocking this state initiates a backward search over \emph{proof obligations}.
An obligation $(q,i)$ asks IC3 to prove that the states described by the formula $q(X)$ are unreachable within $i$ transitions.
Its predecessor query is
\[
F_{i-1}(X)\land T(X,X')\land q(X').
\]
In the left panel of Figure~\ref{fig:ic3-explanation}, a SAT query finds a predecessor $p_{k-1}$ of the bad state in $F_{k-1}$.
\emph{SAT generalization} expands $p_{k-1}$ into a region $q_{k-1}$ whose states can reach the parent obligation, creating $(q_{k-1},k-1)$.
Repeating this process produces earlier obligations, such as $(q_i,i)$ around $p_i$.
These obligations form a backward search tree; the figure shows one branch, with arrows indicating forward transitions.
A branch reaching an initial state yields a counterexample~\cite{PDR}.

If the predecessor query is UNSAT and $q$ excludes initial states, IC3 can block $(q,i)$.
\emph{UNSAT generalization} seeks a lemma $\ell$ satisfying $q\Rightarrow\neg\ell$.
The direct blocking lemma is $\neg q$; a stronger lemma excludes additional states.
To admit $\ell$ at frame $i$, IC3 checks initiation and relative inductiveness:
\[
\mathrm{Init}\Rightarrow\ell,\qquad
F_{i-1}\land\ell\land T\Rightarrow\ell'.
\]
The checked lemma is added to $F_1,\ldots,F_i$, strengthening their approximations and blocking $(q,i)$.
The right panel of Figure~\ref{fig:ic3-explanation} illustrates this step for $(q,k-1)$: the orange region $\neg\ell$ contains $q$ and is excluded by adding $\ell$ to $F_{k-1}$.
Once all bad states have been blocked from the frontier, IC3 extends the frame sequence.
If two adjacent frames $F_i$ and $F_{i+1}$ become equivalent, $F_i$ is an inductive invariant proving the property~\cite{Bradley2011}.

The representation of proof obligations and lemmas determines the available generalizations.
In bit-level IC3, an obligation is typically a cube, a conjunction of Boolean literals, and a blocking lemma is a clause, a disjunction of literals.
Removing literals enlarges the region described by a cube: SAT generalization obtains a broader predecessor region, while UNSAT generalization obtains a stronger blocking lemma by negating the enlarged region~\cite{PDR}.

Word-level IC3 can express lemmas over multi-bit values using predicates such as equality and arithmetic comparisons~\cite{SyGuSAPDR}.
For example, equality between two $n$-bit registers $u$ and $v$ is expressed by the single word-level lemma $u=v$.
Its bit-level representation consists of the following $2n$ clauses:
\[
(\neg u[j]\lor v[j])
\qquad\mbox{and}\qquad
(u[j]\lor\neg v[j]),
\quad 0\le j<n.
\]

%% file: sections/motivation.tex
\section{Motivating Example}
\label{sec:motivation}

Figure~\ref{fig:motivation-controller} compares generalization in Pono-IC3Bits\footnote{Pono-IC3Bits denotes the IC3Bits engine in the Pono model checker~\cite{Pono}, used here as a conventional IC3 reference.} with RTL-guided agent generalization on an information-flow verification example.
Two copies of a controller share input \texttt{a} and reset, while their \texttt{b} operands may differ when \texttt{a} is zero and must otherwise agree.
The assertion requires both copies to produce identical \texttt{finish} signals, checking that the permitted differences in \texttt{b} do not affect completion timing.

\begin{figure}[!htb]
\begin{lstlisting}[style=rtl,basicstyle=\footnotesize\ttfamily]
module controller (
  input wire clk, rst,
  input wire [7:0] a, b,
  output reg finish
);
  localparam IDLE=2'd0, WORK=2'd2, DONE=2'd3;
  reg [1:0] state;
  reg [7:0] count;
  always @(posedge clk)
    if (rst) begin
      state <= IDLE; count <= 0; finish <= 0;
    end else case (state)
      IDLE: begin
        count <= (a == 0) ? 8'd0 : b;
        state <= WORK;
      end
      WORK: if (count == 0) state <= DONE;
            else count <= count - 1'b1;
      DONE: finish <= 1;
    endcase
endmodule

module duv (
  input wire clk, rst,
  input wire [7:0] a, b1, b2
);
  wire finish1, finish2;
  controller copy1 (clk, rst, a, b1, finish1);
  controller copy2 (clk, rst, a, b2, finish2);
  assume property (a == 0 || b1 == b2);
  assert property (finish1 == finish2);
endmodule
\end{lstlisting}
\caption{Two-copy information-flow security verification written in Verilog. Both copies start from the reset
state. 
\texttt{a} and reset are shared, while \texttt{b1} and \texttt{b2}
are constrained by the assumption.}
\label{fig:motivation-controller}
\end{figure}

Write $s_j$ and $f_j$ for the state and finish signals of copy $j\in\{1,2\}$.
Searching backward from unequal finish signals, IC3 obtains the predecessor region
\[
q=\neg f_1\land(s_1=\mathrm{DONE})
  \land\neg f_2\land(s_2=\mathrm{WORK}).
\]
With reset deasserted, the next transition sets $f_1$ and leaves $f_2$ low, reaching a bad state.
The predecessor query against $F_0=\mathrm{Init}$ is UNSAT, showing that $q$ can be blocked at frame~1.

Pono-IC3Bits greedily removes literals while preserving the blocking conditions~\cite{PDR}.
Generalizing $q$ to $s_1[0]=1$ yields the blocking lemma
\[
\ell_{\mathrm{local}}\equiv(s_1[0]=0).
\]
After one transition from initialization, the controller can only be in \texttt{IDLE} or \texttt{WORK}.
The lemma therefore excludes \texttt{DONE} at frame~1, but \texttt{DONE} becomes reachable in later frames.

Given the RTL and $q$, the agent instead proposes
\[
\ell_{\mathrm{sync}}\equiv(s_1=s_2).
\]
The design and input assumption explain this relationship: both counters load zero when \texttt{a} is zero and otherwise load equal \texttt{b} operands.
Their identical updates keep the controllers synchronized.
The lemma excludes $q$ and captures a relationship across execution phases that contributes to the inductive invariant proving equal completion timing.

%% file: sections/framework.tex
\section{Agentic-IC3 Framework}
\label{sec:framework}

\begin{figure}[!tbp]
  \centering
  \includegraphics[width=.85\linewidth]{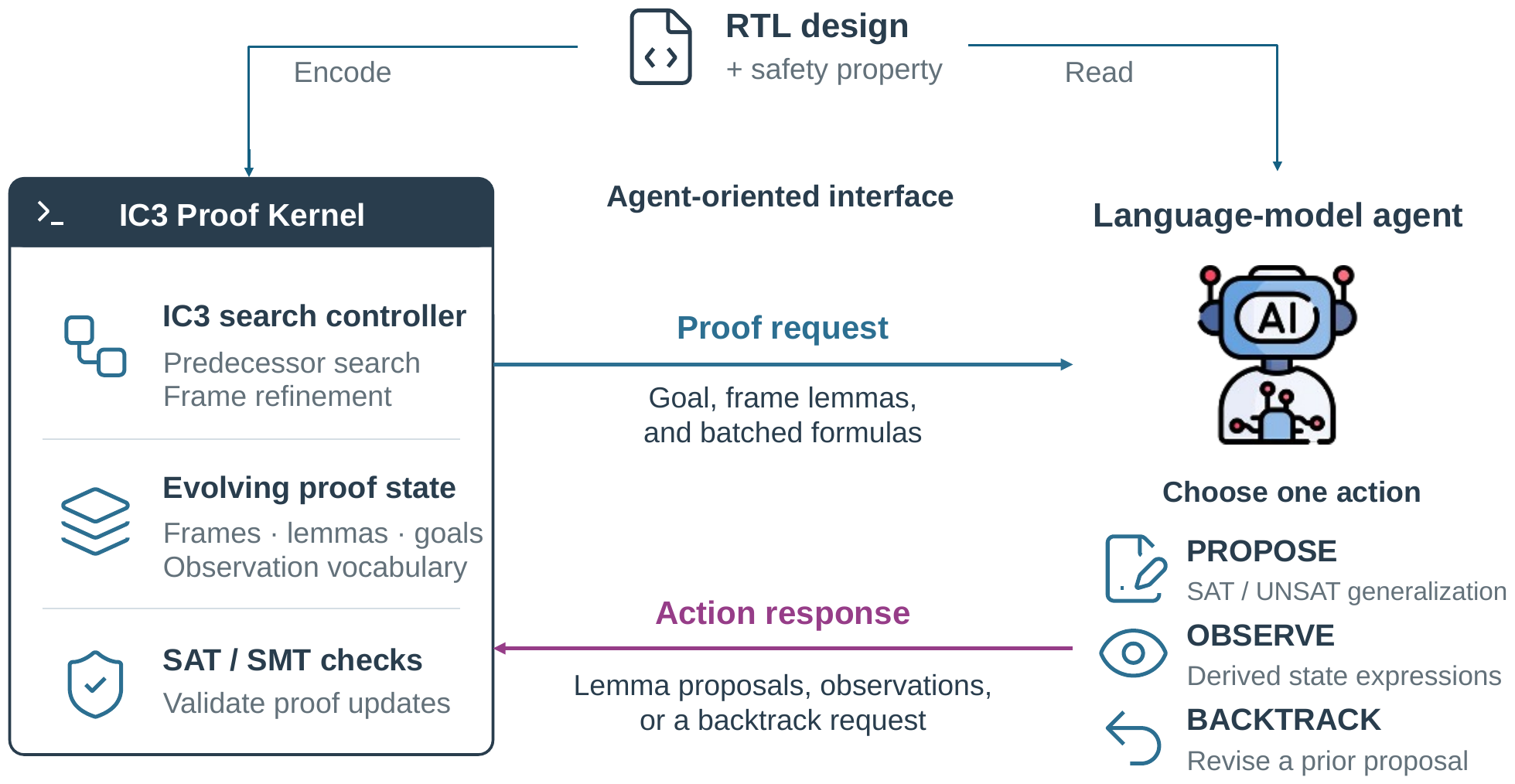}
  \caption{Agentic-IC3 architecture. The kernel sends proof requests to the agent, which uses the RTL and accumulated context to \textsc{Propose}, \textsc{Observe}, or \textsc{Backtrack}. The kernel checks proof updates.}
  \label{fig:agentic-ic3-architecture}
\end{figure}

Agentic-IC3 builds on the separation between proof-search guidance and solver-checked proof steps described in Section~\ref{sec:introduction}.
The agent uses the RTL and evidence from the ongoing search to propose and revise lemmas.
In Figure~\ref{fig:agentic-ic3-architecture}, the IC3 proof kernel comprises a search controller, an evolving proof state, and SAT/SMT checks.
The controller executes IC3 and requests agent guidance at SAT and UNSAT generalization steps.
In response, the agent can \textsc{Propose} lemmas for generalization, \textsc{Observe} derived expressions whose values augment subsequent feedback, or \textsc{Backtrack} to an earlier proof obligation to revise a proposal.
We next describe the interface, the motivation for each action, and its backend support.

\subsection{Design and Agent Interface}
\label{sec:agent-interface}

\paragraph{RTL proof representation.}
Agentic-IC3 uses Pono’s word-level model-checking infrastructure~\cite{Pono} to allow lemmas to directly express relationships inferred from the RTL, such as equality between counters or bounds on arithmetic expressions. The agent reads the RTL, property, and assumptions, and receives the current-state variables and their bit widths. 
The interface presents proof obligations, frame lemmas, and solver feedback using RTL signal names and expressions. 
The agent proposes lemmas using a subset of Verilog expressions, including arithmetic, comparisons, bitwise operations, and concatenation. 
A Verilog-expression frontend parses and type-checks these proposals and compiles them into quantifier-free bit-vector SMT formulas used by the kernel.

\paragraph{Persistent proof session.}
The kernel maintains a persistent proof session in which frames, admitted lemmas, proof obligations, and registered observations remain available across requests.
Agent actions advance or revise this proof state, while the agent uses accumulated interaction context to reason across successive obligations.
A long-lived CLI process retains the solver context and exchanges requests and responses through JSON Lines.

Each proof request identifies the obligation and its frame level $i$, and supplies the active lemmas of $F_{i-1}$.
Depending on the generalization task, the request also contains a batch of predecessor regions or checked blocking lemmas. The backend collects these lemmas before adding them to the frames, allowing the agent to combine low-level facts into more informative design relationships through batch generalization.
Each agent response selects one action: \textsc{Propose}, \textsc{Observe}, or \textsc{Backtrack}.

\subsection{Batch Generalization}
\label{sec:batch-generalization}

Conventional IC3 performs generalization on an individual predecessor state or proof obligation.
Directly replacing these generalization steps with agent calls gives the agent only one example at a time.
The agent may then propose a lemma based on particular register values in that example, without identifying the design relationships that explain other similar states.
IC3Syn addresses this difficulty by presenting multiple states together for generalization~\cite{IC3Syn}.
Following this idea, Agentic-IC3 provides batches at both SAT and UNSAT generalization steps.
SAT batches contain predecessor states and their generalized regions, while UNSAT batches contain blocking lemmas obtained during IC3 search.
The agent examines each batch together with the RTL, frame lemmas, and previous interactions.

We use $b$ for the maximum batch size and $m_{\max}$ for the maximum number of lemmas per proposal.
A proposal contains $\ell_1,\ldots,\ell_m$, where $1\leq m\leq m_{\max}$.

We first describe SAT and UNSAT batch generalization for responses containing lemma proposals.
Both algorithms use \textsc{RequestGeneralization} to query the agent with a proof request $r$ containing the current batch and proof context.
The following subsections extend this interaction with observation and backtracking actions.

\begin{algorithm}[!tb]
  \caption{SAT batch generalization}
  \label{alg:sat-batch}
  \algrenewcommand{\algorithmicindent}{1em}
  \algrenewcommand{\algorithmicrequire}{\textbf{Input:}}
  \begin{algorithmic}[1]
    \Require Proof obligation $(q,i)$, $i>0$.
    \State $E\gets F_{i-1}\land T\land q'$; $\mathcal{B}\gets\emptyset$
    \While{$|\mathcal{B}|<b$ and $\mathrm{SAT}(E)$}\alglabel{line:sat-enumerate}
      \State $p\gets$ predecessor satisfying $E$
      \State $G\gets\mathrm{Generalize}(p)$
      \State $\mathcal{B}\gets\mathcal{B}\cup\{(p,G)\}$
      \State $E\gets E\land\neg G$\alglabel{line:sat-exclude}
    \EndWhile
    \If{$\mathcal{B}=\emptyset$}
      \State \Return to UNSAT generalization
    \EndIf
    \State $r\gets$ SAT request for $(q,i)$ with $\mathcal{B}$ and active frame lemmas
    \State $(\ell_1,\ldots,\ell_m)\gets\Call{RequestGeneralization}{r}$
    \State $L\gets\bigwedge_{j=1}^{m}\ell_j$
    \If{$\mathrm{UNSAT}(F_{i-1}\land T\land q'\land\neg L)$}\alglabel{line:sat-coverage}
      \State Report that no predecessor is excluded
      \State \Return for revision
    \EndIf
    \For{each $\ell_j$ in proposal order}
      \State Check $\ell_j$ with BMC to depth $i-1$\alglabel{line:sat-bmc}
      \If{a reachable violation is found}
        \State \Return counterexample as proposal feedback
      \Else
        \State Recursively block $(\neg\ell_j,i-1)$\alglabel{line:sat-recursive}
      \EndIf
      \If{$\mathrm{UNSAT}(F_{i-1}\land T\land q')$}\alglabel{line:sat-recheck}
        \State \Return to UNSAT generalization
      \EndIf
    \EndFor
    \State Continue with remaining predecessors
  \end{algorithmic}
\end{algorithm}
\noindent\textbf{SAT generalization.}
For a proof obligation $(q,i)$, IC3 must exclude its predecessors from $F_{i-1}$ before it can block $q$ at frame $i$.
Algorithm~\ref{alg:sat-batch} first collects up to $b$ predecessors satisfying $F_{i-1}\land T\land q'$.
Each predecessor $p$ is expanded into a region $G$ using conventional SAT generalization~\cite{PDR}.
The next enumeration query excludes $G$, so subsequent predecessors represent states outside the regions already collected (line~\ref{line:sat-exclude}).
This avoids collecting states already covered by the same conventional generalization.

The agent uses this batch to propose lemmas $\ell_1,\ldots,\ell_m$ for $F_{i-1}$.
Each proposal $\ell_j$ expresses a relationship expected to hold within the first $i-1$ transitions; its negation $\neg\ell_j$ describes the states to exclude.
Different predecessors may require different lemmas to explain their unreachability.
We allow proposals to exclude only some predecessors, so the agent need not identify every relationship in one response.
The backend checks whether any predecessor violates the proposed conjunction $L=\bigwedge_j\ell_j$ by testing the satisfiability of $F_{i-1}\land T\land q'\land\neg L$ (line~\ref{line:sat-coverage}).
If this formula is UNSAT, none of the predecessors would be removed, and the agent is asked to revise its proposals.

Conventional SAT generalization requires every state in the generalized predecessor region to reach $q$ in one transition~\cite{PDR}.
Agentic-IC3 relaxes this requirement, allowing the agent to propose a lemma $\ell_j$ whose negation also includes states that do not lead directly to $q$.
This allows a relationship inferred from the RTL to exclude more unreachable states than the current predecessor query identifies.
However, the proposed relationship may be too strong and exclude reachable states as well.
Before starting recursive blocking, the backend therefore uses bounded model checking (BMC) to check whether an execution from $\mathrm{Init}$, following the original transition relation $T$, can violate $\ell_j$ within $i-1$ transitions (line~\ref{line:sat-bmc}).
A violating execution is returned to the agent for revision, avoiding further backward search for a proposal that cannot hold at its target frame.

If the BMC check is UNSAT, the proposal holds within the checked bound.
The backend then recursively blocks $(\neg\ell_j,i-1)$ to refine the frames through IC3's initiation and relative-inductiveness checks (line~\ref{line:sat-recursive}). 
After these refinements, it repeats the predecessor query for $(q,i)$.
Remaining predecessors lead to further generalization requests; once the query becomes UNSAT, IC3 proceeds to UNSAT generalization to block $q$.

\begin{algorithm}[!tb]
  \caption{UNSAT batch generalization}
  \label{alg:unsat-batch}
  \algrenewcommand{\algorithmicindent}{1em}
  \algrenewcommand{\algorithmicrequire}{\textbf{Input:}}
  \begin{algorithmic}[1]
    \Require Pending batch $\mathcal{B}$ at frame $i$.
    \State Add newly checked lemmas to $\mathcal{B}$\alglabel{line:unsat-collect}
    \If{$|\mathcal{B}|<b$ and collection can continue}\alglabel{line:unsat-threshold}
      \State \Return to IC3 search
    \EndIf
    \While{$\mathcal{B}\ne\emptyset$}
      \State $r\gets$ UNSAT request at frame $i$ with $\mathcal{B}$ and active frame lemmas
      \State $(\ell_1,\ldots,\ell_m)\gets\Call{RequestGeneralization}{r}$
      \For{each $\ell_j$ in proposal order}
        \State $\mathcal{C}\gets\{B\in\mathcal{B}\mid\ell_j\Rightarrow B\}$\alglabel{line:unsat-coverage}
        \State Check $\mathrm{Init}\Rightarrow\ell_j$\alglabel{line:unsat-init}
        \State Check $F_{i-1}\land\ell_j\land T\Rightarrow\ell_j'$\alglabel{line:unsat-induction}
        \If{both checks pass}
          \State Add $\ell_j$ to $F_1,\ldots,F_i$\alglabel{line:unsat-admit}
          \State $\mathcal{B}\gets\mathcal{B}\setminus\mathcal{C}$\alglabel{line:unsat-remove}
        \Else
          \State \Return counterexample as proposal feedback
        \EndIf
      \EndFor
    \EndWhile
  \end{algorithmic}
\end{algorithm}
\noindent\textbf{UNSAT generalization.}
Conventional IC3 blocks an obligation by deriving a lemma and adding it to the frames.
Several such lemmas may express separate constraints that can be generalized into a simpler or stronger design relationship.
To allow the agent to consider these constraints together, Agentic-IC3 collects checked blocking lemmas in a batch $\mathcal{B}$ before adding them to the frames.
These lemmas may result from blocking individual obligations or from recursively proving lemmas proposed during SAT generalization.

Algorithm~\ref{alg:unsat-batch} requests agent generalization when $b$ lemmas have accumulated, or earlier when the collected lemmas are needed to continue the current search.
The agent examines the batch and the RTL to propose lemmas that replace one or more of the collected constraints.
A proposal $\ell$ can replace a batch lemma $B_j$ when $\ell\Rightarrow B_j$.
Thus, $\neg\ell$ contains $\neg B_j$: the proposal blocks at least the states excluded by $B_j$.

The backend checks initiation, $\mathrm{Init}\Rightarrow\ell$, and relative inductiveness, $F_{i-1}\land\ell\land T\Rightarrow\ell'$.
If both checks pass, it adds $\ell$ to $F_1,\ldots,F_i$ and removes the batch lemmas implied by $\ell$ (lines~\ref{line:unsat-admit}--\ref{line:unsat-remove}).
Otherwise, it returns the counterexample from the failed check for revision.
Generalization continues for the remaining batch lemmas until each is covered by an admitted lemma.

\subsection{Derived Observation Vocabulary}
\label{sec:observation-vocabulary}

Batch generalization allows the agent to compare several states, but useful relationships may not be apparent from individual register values.
Identifying these relationships requires the agent to reason about combinations of signals and compare the resulting values across states.
This requires additional reasoning, and useful relationships may be overlooked.
To provide more informative feedback, Agentic-IC3 allows the agent to introduce named observations whose values are included in subsequent proof requests.

Consider the two four-bit counters \texttt{cnt1} and \texttt{cnt2} and the control signal \texttt{active} in Table~\ref{tab:observation-example}.
Neither counter has a fixed value when \texttt{active} is asserted.
Their sum, however, is $8$ in each of these states.
The agent can introduce an \emph{observation} \texttt{obs.total = cnt1 + cnt2}, adding the last column to the request.
This makes the conditional relationship directly visible: \texttt{obs.total} is $8$ when \texttt{active} is asserted and varies in the other displayed states.
The agent can then propose $\texttt{active}\Rightarrow\texttt{obs.total == 4'd8}$ for generalization.

\begin{table}[t]
  \centering
  \small
  \caption{A derived observation makes a conditional relationship between two counters explicit.}
  \label{tab:observation-example}
  \setlength{\tabcolsep}{6pt}
  \begin{tabular}{@{}cccc@{}}
    \toprule
    \texttt{active} & \texttt{cnt1} & \texttt{cnt2}
      & \texttt{obs.total} \\
    \midrule
    1 & 1 & 7 & \textbf{8} \\
    1 & 5 & 3 & \textbf{8} \\
    1 & 6 & 2 & \textbf{8} \\
    0 & 2 & 3 & 5 \\
    0 & 4 & 7 & 11 \\
    \bottomrule
  \end{tabular}
\end{table}

In \textsc{RequestGeneralization}$(r)$, the agent may respond to the proof request $r$ with $\textsc{Observe}(e)$ instead of a lemma proposal.
The backend checks the expression $e$ and registers its name, definition, and bit width.
It then augments $r$ with the observation values and queries the agent again.
The agent may introduce further observations before returning lemmas.
Registered observations remain available throughout the session and are evaluated on states returned in later requests and counterexamples.
The agent can thus compare the same derived value across successive proof obligations and reuse the observation name to simplify lemma expressions.

\subsection{Backtracking and Proof Revision}
\label{sec:proof-backtracking}

UNSAT generalization adds a proposed lemma to the frames once it passes the initiation and relative-inductiveness checks.
SAT generalization instead creates a proof obligation that the backend tries to block recursively.
The choice of a SAT proposal therefore determines which obligations the subsequent search must prove.
Even if the proposal holds within the current frame bound, proving it through recursive blocking may require many intermediate obligations.

IC3 gains efficiency from local SAT/SMT queries involving only one transition, without unrolling the transition relation~\cite{Bradley2011}.
However, the resulting feedback describes only the current proof step, making it difficult for the agent to anticipate the recursive obligations needed to establish a proposed lemma.
As backward search continues, the agent can relate successive predecessor states and learned lemmas to the RTL and develop a better understanding of the design behavior.
It may then recognize that an earlier proposal describes only a local fact or that another relationship offers a more effective direction for the proof.
Agentic-IC3 therefore allows the agent to backtrack and replace the earlier proposal before completing its proof attempt.

For a proposal $\ell$ made while blocking $(q,i)$, Algorithm~\ref{alg:sat-batch} recursively attempts to block $(\neg\ell,i-1)$ (line~\ref{line:sat-recursive}).
When \textsc{RequestGeneralization}$(r)$ is invoked for SAT generalization of this obligation, the agent may return \textsc{Backtrack} instead of lemmas.
This response ends the recursive attempt to prove $\ell$ and returns control to the generalization step for $(q,i)$, skipping the remaining lemmas from the same proposal response.
The backend repeats the query $F_{i-1}\land T\land q'$ using the retained frame lemmas (line~\ref{line:sat-recheck}).
If predecessors remain, it collects a new batch and calls \textsc{RequestGeneralization} with an updated request for $(q,i)$.
Admitted lemmas remain in their validated frames and can prune subsequent search, since each has passed its own proof checks independently of the abandoned proposal.

%% file: sections/evaluation.tex
\section{Evaluation}
\label{sec:evaluation}

We compare agent-guided proof search with conventional and ML-assisted IC3 for hardware verification.

We implement Agentic-IC3 in Python using Pono's word-level model-checking infrastructure~\cite{Pono}.
Yosys translates the RTL into a word-level transition system in BTOR2 format~\cite{Yosys}, and Bitwuzla handles the backend SMT queries~\cite{Bitwuzla}.
We use Codex CLI version 0.153.0 as the agent framework, with GPT-5.6-Sol and medium reasoning effort.
The full Agentic-IC3 configuration uses a maximum batch size of $b=5$ for both SAT and UNSAT generalization and allows at most $m_{\max}=5$ lemmas per proposal.

Our evaluation comprises 14 benchmarks spanning security information-flow verification and functional verification of communication protocols, processors, and functional units.
The suite includes cache and arithmetic-unit information-flow checks adapted from SecIC3~\cite{SecIC3}; FIFO, arithmetic-loop, and arbitration benchmarks from the hard benchmark set of Large Lemma Miners~\cite{LargeLemmaMiners}; and processor-verification benchmarks for NERV, PicoRV32, and SERV from CIll~\cite{CIll}.
We also include a buffer-overflow check for the BSG communication protocol, adapted from the ILA models used by Lu et al.~\cite{SoCProtocol}.
We primarily select cases that are difficult for conventional IC3 engines, while retaining a small number of easier cases to assess the overhead of agent-guided verification.

Each run starts with a fresh agent conversation and persistent IC3 session.
The full-system runs use the same task prompt and interface documentation across benchmarks.
The agent has read-only access to the RTL, documentation, and proof history, and writable space for temporary files.
It interacts with the kernel through the shell; backend source code and internal state files remain outside the sandbox.
Shell network access and web search are disabled.

We compare Agentic-IC3 with Pono-IC3Bits~\cite{Pono}, rIC3~\cite{rIC3}, and $\mathcal{A}$-IC3~\cite{AIC3}.
Pono-IC3Bits provides a conventional IC3 baseline using the same model-checking infrastructure.
rIC3 provides a state-of-the-art bit-level IC3 baseline, while $\mathcal{A}$-IC3 provides an ML-assisted baseline that adaptively selects generalization strategies.
All experiments run on AMD EPYC 7532 processors, with one CPU core and 16~GiB of memory allocated per run and a verification timeout of 3,600 seconds.
We report wall-clock runtimes and additionally distinguish the time spent awaiting agent responses from work performed by the Agentic-IC3 backend.

\subsection{Comparison with IC3 Baselines}
\label{sec:baseline-comparison}

\input{Tables/tab-eval-runtime-breakdown}

Table~\ref{tab:main-results} summarizes the verification results.
Agentic-IC3 solves 10 of the 14 benchmarks within one hour, whereas Pono-IC3Bits times out on all 14.
Agentic-IC3 thus solves additional cases using the same model-checking infrastructure.
Against the other baselines, rIC3 and $\mathcal{A}$-IC3 each solve five cases.
Agentic-IC3 solves every case solved by either baseline and four additional cases: \texttt{lrg\_arb\_lrg\_16\_ebmc}, \texttt{fp\_divider}, \texttt{nerv\_causal}, and \texttt{picorv32\_pc\_bwd}.

Agentic-IC3 also improves runtime on \texttt{cache}, \texttt{fifo\_vis}, and \texttt{fp\_adder} compared with both baselines.
Agent interaction introduces overhead, however: on \texttt{multiplier}, rIC3 and $\mathcal{A}$-IC3 finish in approximately three seconds, compared with 21.23 seconds for Agentic-IC3.
Across the ten completed runs, time spent awaiting agent responses accounts for 87\% of the aggregate runtime.
This suggests that reducing agent-response latency could substantially improve end-to-end performance.
One potential direction is to fine-tune smaller models on recorded IC3 proof traces, aiming to retain effective generalization and action selection with faster inference.

\subsection{Ablation Study of Individual Mechanisms}
\label{sec:ablation}

We evaluate the contribution of batch generalization, observations, and backtracking by disabling each mechanism separately on the ten benchmarks solved by the full system with all mechanisms enabled.
Without batching, each SAT or UNSAT request contains one item, and the agent is asked to propose one lemma per response.
The other two configurations disable the corresponding action, with the interface documentation adjusted accordingly.
Table~\ref{tab:ablation-results} reports the outcome and runtime of each run, together with the numbers of generalizations, backtracks, and registered observations.
We examine both proof completion and generalization counts, since the reasoning required for individual agent responses varies and runtime alone does not describe the resulting search.

\input{Tables/tab-eval-ablation}

\paragraph{Batch generalization.}
The full system requires fewer generalizations than the no-batching configuration on all ten benchmarks.
The total decreases from 515 to 137, a reduction of 73\%, with the largest reductions on \texttt{fifo\_vis} (173 to 7) and \texttt{cache} (22 to 2).
Batching lets the agent generalize several predecessors or blocking lemmas together, reducing the number of generalization calls.
Both configurations solve all ten benchmarks, and batching also reduces runtime on nine, with speedups of $3.3\times$ on \texttt{fp\_divider} and $2.6\times$ on \texttt{fp\_adder}.

\paragraph{Observations.}
The full system solves one more benchmark than the configuration without observations.
On \texttt{gulwani\_fig1a\_2\_ebmc}, it completes the proof in 175.73 seconds using two observations and seven generalizations, whereas the no-observations run makes 41 generalizations and times out.
On \texttt{cache} and \texttt{fifo\_vis}, the full system also requires half as many generalizations: two rather than four, and seven rather than fourteen, respectively.
On \texttt{fifo\_vis}, observations reduce the number of generalizations even though the two runs take similar time.
The trajectory in Section~\ref{sec:trajectory-decisions} shows how an occupancy observation helps relate two different FIFO representations.
The run without observations is faster on \texttt{bsg\_link}, but this difference may reflect variation between runs; repeated runs are needed to determine whether it is consistent.

\paragraph{Backtracking.}
The full-system runs for \texttt{fp\_adder} and \texttt{nerv\_causal} each backtrack once.
Compared with the runs without backtracking, their generalization counts decrease from 67 to 23 and from 43 to 11, with runtime speedups of $2.3\times$ and $10.5\times$, respectively.
Although both configurations solve all ten benchmarks, these cases show substantial differences in the amount of generalization needed to complete the proof.
The adder trajectory in Section~\ref{sec:trajectory-decisions} explains the role of backtracking: recursive feedback reveals that an earlier proposal requires operand relationships before the corresponding registers are updated.
Backtracking lets the agent replace that proposal with relationships restricted to the relevant control states, while retaining previously admitted lemmas.

\subsection{Proof-Search Trajectory Analysis}
\label{sec:trajectory-analysis}

We examine recorded proof-search trajectories to show how the agent uses design information and solver feedback to propose lemmas, introduce observations, and revise earlier proposals.
We also examine difficulties encountered in both completed and timed-out runs.

\subsubsection{Agent Decisions}
\label{sec:trajectory-decisions}

\paragraph{Generalization from design semantics.}
The \texttt{multiplier} benchmark checks whether two multiplier instances produce their completion signals simultaneously under an input constraint.
For instance $j\in\{1,2\}$, let $a_j$ and $b_j$ denote its stored operands and $\mathit{busy}_j$ its busy flag.
Each instance asserts its completion signal on the next cycle when it is busy on the current cycle and either stored operand is zero.
The initial predecessor batch contains states in which the instances disagree on these conditions.
The agent proposes three relationships:
\[
a_1=a_2,\qquad
(a_1=0)\lor(b_1=b_2),\qquad
\mathit{busy}_1=\mathit{busy}_2.
\]
These relationships capture the completion logic: when the shared operand is zero, the other operands need not agree; otherwise, their equality ensures that both instances detect zero simultaneously.
The backend admits these three relationships and completes the proof.

\paragraph{Observations connecting different representations.}
The \texttt{fifo\_vis} benchmark compares a shift-register FIFO (\texttt{sr}) with a ring-buffer FIFO (\texttt{rb}), each containing sixteen entries.
In the shift-register FIFO, the tail index (\texttt{sr.tail}) identifies the oldest entry, and the empty flag (\texttt{sr.empty}) indicates whether any entries are present.
When nonempty, this FIFO contains $\mathtt{sr.tail}+1$ entries.
The ring-buffer pointers \texttt{rb.head} and \texttt{rb.tail} identify the next insertion location and oldest entry, respectively.

The agent introduces the four-bit observation $\mathtt{obs.occupancy}=\mathtt{rb.head}-\mathtt{rb.tail}$, representing ring-buffer occupancy modulo 16.
Subsequent feedback includes this value alongside the original state variables.
For example, one predecessor has $\mathtt{rb.head}=5$ and $\mathtt{rb.tail}=4$, giving $\mathtt{obs.occupancy}=1$, while the nonempty shift-register FIFO has $\mathtt{sr.tail}=7$, representing eight entries.
A pop therefore exposes inconsistent empty flags.

The agent proposes a lemma relating the occupancy representations: its nonempty case requires $\mathtt{obs.occupancy}=\mathtt{sr.tail}+1$, with separate conditions for initialization and the empty case.
All arithmetic in these expressions is four-bit arithmetic modulo 16; the empty flags distinguish an empty FIFO from a full FIFO.
The backend admits the lemma expressed using this observation.

\paragraph{Backtracking after requiring equality too early.}
The \texttt{fp\_adder} benchmark checks whether two adder instances assert their completion signals together.
For instance $j\in\{1,2\}$, let $s_j$, $a_j$, and $b_j$ denote the RTL registers \texttt{state}, \texttt{a}, and \texttt{b}, respectively.
Each instance loads its operands in separate cycles, then decodes them.
The following simplified RTL shows the relevant clocked updates; \texttt{take\_a} and \texttt{take\_b} abbreviate the input handshakes, and unrelated assignments are omitted.

\begin{figure}[th]
\begin{lstlisting}[style=rtl,basicstyle=\footnotesize\ttfamily,aboveskip=4pt,belowskip=4pt]
case (state)
  0: if (take_a) begin a <= input_a; state <= 1; end
  1: if (take_b) begin b <= input_b; state <= 2; end
  2: begin
       b_e <= b[30:23] - 127; // also update b_m, b_s
       state <= 3;
     end
endcase
\end{lstlisting}
\caption{Code snippet for the \texttt{fp\_adder} state transitions.}
\end{figure}
On entering state 1, $a_j$ has been updated but $b_j$ still holds its old value.
The new $b_j$ is available in state 2, and its decoded exponent, mantissa, and sign (\texttt{b\_e}, \texttt{b\_m}, \texttt{b\_s}) are available in state 3.

While proving $s_1=s_2$, the agent initially requires $b_1=b_2$ and equality of the corresponding decoded fields whenever $a_1\neq0$.
The backend admits this relationship at an early frame.
When the agent later tries to establish it at a higher frame, recursive search returns a transition from state 0 to state 1 that loads a nonzero $a_1$ while leaving unequal $b_1$ and $b_2$ and their decoded fields unchanged.
The proposal therefore requires equality before those registers have been updated.

The agent backtracks to the proof of $s_1=s_2$ and replaces the proposal with two lemmas.
The first requires $b_1=b_2$ when $s_1\geq2$ and $a_1\neq0$.
The second requires equality of the decoded fields when $3\leq s_1\leq10$ and $a_1\neq0$.
The backend admits the revised lemmas, and the subsequent search completes the proof.

\subsubsection{Limitations and Extensions}
\label{sec:trajectory-limitations}

\paragraph{Quantified lemmas to reduce enumeration.}
Relationships that hold uniformly across indexed state components can require many explicit cases in the current lemma language.
Allowing quantified lemmas could let the agent express such relationships over all relevant indices in a single formula.
The \texttt{fifo\_vis} trajectory illustrates this opportunity: one admitted lemma uses sixteen disjuncts to relate a shift-register word to the corresponding ring-buffer word under each possible circular address offset, and similar lemmas cover other words.
Symbolic array indexing could remove the explicit address cases, while universal quantification could express the correspondence across all valid entries.
Together, these extensions could make proposals more concise and reduce repeated reasoning about individual indices.
They would also require backend support for checking the richer formulas; a more compact representation does not by itself guarantee faster verification.

\paragraph{Stateful observations to capture temporal relationships.}
Some useful proof relationships connect values from different execution stages, while the current observation mechanism only defines expressions over the existing state.
Extending observations with persistent state could retain selected past values or summarize execution history, making these relationships available to lemma generation.
The timed-out \texttt{picorv32\_insn\_sub} trajectory illustrates this limitation: proposals repeatedly confuse values belonging to different instructions or assume that a result has reached the retirement record before the corresponding update.
A stateful observation, such as a shadow register or history monitor, could retain an instruction's operands and destination through retirement, providing a stable reference for expressing the intended correspondence.
Such an extension would require explicit initialization and update rules, with auxiliary state that records execution history without restricting the original design's behaviors.
The backend would then check proposed lemmas over the augmented transition system.

%% file: Tables/tab-eval-runtime-breakdown.tex
\begin{table}[ht]
\centering
\caption{End-to-end runtimes in seconds. Agent measures time awaiting agent responses; Backend measures verification work. The remainder is framework and orchestration overhead. TO: one-hour timeout; OOM: out of memory.}
\label{tab:main-results}
\setlength{\tabcolsep}{3pt}
\begin{tabular}{lrrrrrr}
\toprule
\multirow{2}{*}{\textbf{Test case}}
& \multirow{2}{*}{\shortstack{\textbf{Pono-IC3Bits}}}
& \multirow{2}{*}{\textbf{rIC3}}
& \multirow{2}{*}{\textbf{$\mathcal{A}$-IC3}}
& \multicolumn{3}{c}{\textbf{Agentic-IC3}} \\
\cmidrule(l){5-7}
& & & & \textbf{Total} & \textbf{Agent} & \textbf{Backend} \\
\midrule
\texttt{bsg\_link}
& TO & TO & 407.25 & 360.57 & 341.00 & 2.66 \\
\texttt{cache}
& TO & 199.17 & 135.70 & 82.40 & 62.74 & 5.33 \\
\texttt{fifo\_vis}
& TO & 239.80 & 274.79 & 138.35 & 125.51 & 2.30 \\
\texttt{gulwani\_fig1a\_2\_ebmc}
& TO & 51.43 & TO & 175.73 & 129.49 & 33.90 \\
\texttt{lrg\_arb\_lrg\_16\_ebmc}
& TO & TO & TO & 55.47 & 37.64 & 8.73 \\
\texttt{multiplier}
& TO & 2.93 & 3.14 & 21.23 & 13.17 & 0.22 \\
\texttt{fp\_adder}
& TO & 530.81 & 209.53 & 153.22 & 136.54 & 5.89 \\
\texttt{fp\_divider}
& TO & TO & TO & 512.71 & 463.79 & 38.61 \\
\texttt{nerv\_causal}
& TO & TO & OOM & 162.34 & 142.83 & 9.86 \\
\texttt{picorv32\_insn\_mul}
& TO & TO & TO & TO & -- & -- \\
\texttt{picorv32\_insn\_sub}
& TO & TO & TO & TO & -- & -- \\
\texttt{picorv32\_pc\_bwd}
& TO & TO & TO & 42.10 & 32.90 & 1.11 \\
\texttt{picorv32\_reg}
& TO & TO & TO & TO & -- & -- \\
\texttt{serv\_insn\_add}
& TO & TO & TO & TO & -- & -- \\
\midrule
\textbf{Solved}
& \textbf{0/14} & \textbf{5/14} & \textbf{5/14}
& \multicolumn{3}{c}{\textbf{10/14}} \\
\bottomrule
\end{tabular}
\end{table}

%% file: Tables/tab-eval-ablation.tex
\begin{table*}[t]
\centering
\caption{Ablation results on the ten benchmarks solved by the full
system. Gen.\ counts lemma-proposal responses; BT.\ counts backtracks;
Obs.\ counts registered observation definitions. Time is end-to-end
wall time in seconds.}
\label{tab:ablation-results}
\setlength{\tabcolsep}{3pt}
\resizebox{\textwidth}{!}{%
\begin{tabular}{l*{4}{rrrr}}
\toprule
\multirow{2}{*}{\textbf{Test case}}
& \multicolumn{4}{c}{\textbf{Full}}
& \multicolumn{4}{c}{\textbf{No batching}}
& \multicolumn{4}{c}{\textbf{No observations}}
& \multicolumn{4}{c}{\textbf{No backtracking}} \\
\cmidrule(lr){2-5}
\cmidrule(lr){6-9}
\cmidrule(lr){10-13}
\cmidrule(lr){14-17}
& Gen. & BT. & Obs. & Time
& Gen. & BT. & Obs. & Time
& Gen. & BT. & Obs. & Time
& Gen. & BT. & Obs. & Time \\
\midrule
\texttt{bsg\_link}
& 16 & 0 & 6 & 360.57
& 28 & 4 & 7 & 384.04
& 10 & 0 & 0 & 214.41
& 15 & 0 & 7 & 323.58 \\

\texttt{cache}
& 2 & 0 & 6 & 82.40
& 22 & 0 & 13 & 165.42
& 4 & 0 & 0 & 67.66
& 16 & 0 & 0 & 168.17 \\

\texttt{fifo\_vis}
& 7 & 0 & 33 & 138.35
& 173 & 0 & 18 & 445.94
& 14 & 0 & 0 & 132.22
& 46 & 0 & 97 & 543.00 \\

\texttt{gulwani\_fig1a\_2\_ebmc}
& 7 & 0 & 2 & 175.73
& 37 & 0 & 1 & 302.38
& 41 & 0 & 0 & TO
& 10 & 0 & 2 & 358.79 \\

\texttt{lrg\_arb\_lrg\_16\_ebmc}
& 5 & 0 & 0 & 55.47
& 12 & 0 & 0 & 91.81
& 6 & 0 & 0 & 67.43
& 8 & 0 & 0 & 60.01 \\

\texttt{multiplier}
& 2 & 0 & 0 & 21.23
& 4 & 0 & 0 & 30.43
& 4 & 0 & 0 & 47.48
& 2 & 0 & 0 & 26.00 \\

\texttt{fp\_adder}
& 23 & 1 & 0 & 153.22
& 79 & 1 & 2 & 397.62
& 42 & 0 & 0 & 255.46
& 67 & 0 & 0 & 356.16 \\

\texttt{fp\_divider}
& 62 & 0 & 0 & 512.71
& 141 & 1 & 0 & 1691.69
& 109 & 3 & 0 & 710.27
& 78 & 0 & 6 & 440.06 \\

\texttt{nerv\_causal}
& 11 & 1 & 0 & 162.34
& 16 & 1 & 0 & 149.61
& 20 & 4 & 0 & 321.98
& 43 & 0 & 0 & 1704.11 \\

\texttt{picorv32\_pc\_bwd}
& 2 & 0 & 0 & 42.10
& 3 & 0 & 1 & 46.83
& 2 & 0 & 0 & 33.32
& 4 & 0 & 0 & 53.17 \\
\midrule
\textbf{Solved}
& \multicolumn{4}{c}{\textbf{10/10}}
& \multicolumn{4}{c}{\textbf{10/10}}
& \multicolumn{4}{c}{\textbf{9/10}}
& \multicolumn{4}{c}{\textbf{10/10}} \\
\bottomrule
\end{tabular}%
}
\end{table*}

%% file: sections/discussion.tex
\section{Discussion}
\label{sec:discussion}

\paragraph{Combining established forms of proof guidance.}
Agentic-IC3 draws on established ideas in model checking, including synthesis-based lemma generation~\cite{SyGuSAPDR}, extensions to the proof vocabulary~\cite{PDRER}, and global guidance based on previously learned lemmas~\cite{GlobalGuidance}.
Applying these ideas effectively requires deciding which relationships to express, what evidence to examine, and when to revise the search.
Agentic-IC3 provides a common interface through which a language-model agent can combine these forms of guidance according to the evolving proof context.
The backend maintains the algorithm's proof conditions, while the agent selects and applies the available actions.
This suggests a research direction centered on identifying useful proof-search actions and providing informative feedback that enables an agent to use them together.

\paragraph{Understanding and debugging proof search.}
The interaction trace exposes proposed relationships, solver feedback, and the agent's explanations for revising earlier decisions.
These records help connect proof-search behavior to the RTL.
For example, the floating point adder trajectory in Section~\ref{sec:trajectory-decisions} reveals that operand equality and equality of decoded fields become relevant at different control states.
Examining such trajectories can reveal missing observations, repeated unproductive proposals, and opportunities for additional interface actions.
This makes the framework useful for investigating how proof guidance succeeds or fails and for refining the information exposed to the agent.
The resulting insights could also inform conventional IC3 heuristics, such as recognizing control conditions under which a proposed relationship becomes useful.

\paragraph{Learning specialized proof-search policies.}
Recorded proof-search traces provide a possible basis for fine-tuning smaller models for lemma generation and action selection.
Training examples could associate proof contexts and agent decisions with solver feedback and subsequent proof progress.
Since a valid lemma may contribute little to the final invariant, useful training signals should account for the progress and cost of the resulting search.
Such training could improve decisions about when to introduce observations or revise proposals, in addition to reducing response latency.
The same separation between guidance and checked proof steps could extend to other algorithms.
In interpolation-based model checking, for example, interpolant strength affects the resulting approximations and verification performance~\cite{InterpolantStrength}.
An agent could guide interpolant construction or selection, while the backend enforces the interpolation conditions and any additional requirements of the model-checking algorithm.

\paragraph{Alternative integration and global reflection.}
Our framework requests agent guidance at designated generalization steps.
An alternative could allow conventional IC3 heuristics to run until repeated obligations or limited frame progress trigger agent intervention.
Such selective intervention could retain access to the internal proof state while reducing the frequency of agent calls.
A complementary direction is to give the agent an explicit opportunity to review broader proof progress.
Global guidance in IC3-style algorithms already demonstrates the value of considering learned lemmas collectively to address limitations of local reasoning~\cite{GlobalGuidance}.
Drawing on agent reflection~\cite{Reflexion} and multi-agent frameworks~\cite{AutoGen}, a periodic review or a separate reviewing agent could examine admitted lemmas, unsuccessful proposals, and frame progress to suggest revisions to the proof-search policy.
Providing this broader context could help identify patterns that are difficult to recognize while concentrating on one proof obligation at a time.
The benefit of these designs would need to be evaluated against their additional reasoning and coordination costs.

%% file: sections/related-work.tex
\section{Related Work}

Section~\ref{sec:introduction} reviews conventional and ML-based IC3 generalization.
IC3Syn integrates LLM-generated blocking clauses into an IC3 controller for distributed protocols and identifies an agent-based workflow as future work~\cite{IC3Syn}.
Agentic-IC3 provides such an interaction framework for RTL hardware verification and extends the role of learned guidance beyond lemma generation,
allowing an agent to introduce observations and revise proof obligations within a persistent IC3 session.

Large Lemma Miners and CIll generate helper invariants using design information and verification feedback~\cite{LargeLemmaMiners,CIll}.
Agentic-IC3 guides the model checker's internal proof search and could serve as a verification backend for these frameworks.

%% file: sections/conclusion.tex
\section{Conclusion}

Agentic-IC3 integrates a language-model agent into a persistent IC3 proof session.
Using RTL information and solver feedback, the agent guides word-level generalization, introduces observations, and backtracks to revise proposals.
The backend maintains soundness through solver-checked proof updates.
Agentic-IC3 solves 10 of 14 benchmarks, including four unsolved by all three baselines.
The ablation study shows fewer generalization interactions with batching and benefits from observations and backtracking on selected proof searches.
Future work will examine variability across repeated runs, explore additional agent actions, and extend the framework to other model-checking algorithms.